\documentclass[a4paper,fleqn,usenatbib,useAMS]{mnras}

\usepackage{graphicx}	% Including figure files
\usepackage{amsmath}	% Advanced maths commands
\usepackage{amssymb}	% Extra maths symbols
\usepackage{multicol}        % Multi-column entries in tables
\usepackage{bm}		% Bold maths symbols, including upright Greek
\usepackage{pdflscape}	% Landscape pages

\usepackage[T1]{fontenc}
\usepackage{ae,aecompl}

\usepackage{newtxtext,newtxmath}
\title[Spectral variability in Mrk 530]{Exploring the spectral variability of the Seyfert 1.5 galaxy Markarian 530 with \textit{Suzaku}}

\author[H. J. S. Ehler et al.]{
H. J. S. Ehler,\thanks{E-mail: hehler@ap.smu.ca}
A. G. Gonzalez, and L. C. Gallo
\\
Department of Astronomy and Physics, Saint Mary's University, 923 Robie St, Halifax, NS B3H 3C3, Canada}

\date{Last updated 2015 May 22; in original form 2013 September 5}

\pubyear{2018}

\begin{document}
\label{firstpage}
\pagerange{\pageref{firstpage}--\pageref{lastpage}}
\maketitle

\begin{abstract}
A 2012 \textit{Suzaku} observation of the Seyfert 1.5 galaxy Markarian 530 was analysed and found to exhibit two distinct modes of variability, which were found to be independent from one another. Firstly, the spectrum undergoes a smooth transition from a soft to a hard spectrum. Secondly, the spectrum displays more rapid variability seemingly confined to a very narrow energy band ($\sim1-3$~keV). Three physical models (blurred reflection, partial covering, and soft Comptonisation) were explored to characterise the average spectrum of the observation as well as the spectral state change. All three models were found to fit the average spectrum and the spectral changes equally well. The more rapid variability appears as two cycles of a sinusoidal function, but we cannot attribute this to periodic variability. The Fe~K$\alpha$ band exhibits a narrow $6.4$~keV emission line consistent with an origin from the distant torus.  In addition, features blueward of the neutral iron line are consistent with emission from He-like and H-like iron that could be originating from the highly ionised layer of the torus, but a broad Gaussian profile at $\sim6.7$ keV also fits the spectrum well.
\end{abstract}

\begin{keywords}
galaxies: active -- galaxies: individual: Mrk 530 -- X-ray: galaxies -- galaxies: nuclei
\end{keywords}

%%%%%%%%%%%%%%%%%%%%%%%%%%%%%%%%%%%%%%%%%%%%%%%%%%

%%%%%%%%%%%%%%%%% BODY OF PAPER %%%%%%%%%%%%%%%%%%

%%%%%%%%%%%%%%%%%%%%%%%%%
\section{Introduction} %%
%%%%%%%%%%%%%%%%%%%%%%%%%
Active galactic nuclei (AGN) harbour supermassive black holes (SMBHs) at their centres surrounded by an accretion disc and dusty torus, with both being illuminated by some central X-ray source. AGN are responsible for producing extreme environments and are some of most luminous astrophysical sources in the Universe. The classification of these objects dates back to \cite{Seyfert1943}. At present the AGN unification model postulates that AGN classification depends primarily on viewing angle of the centre-most region \citep{Antonucci1993,UrryPadovani1995,Netzer2015}. For example, Seyfert 1 galaxies offer a relatively clean and unobstructed view of the central engine whereas Seyfert 2 galaxies are characterised by highly absorbed spectra due to our viewing angle intersecting the dusty, optically thick torus. A number of classifications between these extremes exist describing sources displaying characteristics akin to both Seyfert 1 and 2 type objects. These sources are classified as intermediate-type Seyferts (e.g. Seyfert 1.2, 1.5, 1.8).

Common to the basic unification model of AGN are the sites of X-ray emission that contribute to the overall shape of their X-ray spectrum, though these components are most easily studied in Seyfert 1 type objects due to the lack of strong absorption. Studies of such sources have revealed the primary X-ray emission source to be a hot, optically thin cloud of electrons (e.g. corona) that resides somewhere near the black hole, likely at some height above the accretion disc. The exact location and shape of this corona is a field of ongoing research and possible configurations investigated include point-like sources (e.g. \citealt{WilkinsFabian2012}), as well as extended (e.g. \citealt{Gonzalez+2017}), collimated (e.g. \citealt{WilkinsGallo2015a,Gonzalez+2017}), and patchy (e.g. \citealt{WilkinsGallo2015b}) structures. The X-ray emission from this corona takes the form of a power law with photon index $\Gamma\sim1.8$ roughly describing most AGN spectra. However, in most sources a simple power law fit to the spectrum is insufficient to describe the spectral shape and a soft excess below $\sim1$ keV is often observed. 

The soft excess component may be the result of relativistic reflection off of the inner accretion disc (e.g. \citealt{Fabian+1989, GeorgeFabian1991,RossFabian2005}). X-ray emission from the hot corona will go on to illuminate both the accretion disc and dusty torus, producing a reflection spectrum for each component. 

The reflection from the dusty torus shows the material to be  neutral (cold) and distant matter as evidenced by the narrow and neutral Fe~K$\alpha$ emission lines. Reflection off  the inner accretion disc, however, leads to relativistically broadened emission features due to the general relativistic effects experienced by the material being strongly affected by the gravitational field of the black hole (e.g. \citealt{Fabian+1989,Laor1991}). The forest of emission lines in the $<1$ keV band will thus become broadened and blurred into a smooth, featureless hump, therefore accounting for the soft excess component.

An alternative explanation to this soft excess describes a second X-ray corona comprising the innermost region of the accretion disc \citep{Magdziarz+1998}. Here the cooler, more optically thick corona acts as a site for soft Comptonisation of seed UV photons from the accretion disc. In this scenario the soft excess is accounted for by this soft Comptonising corona with the power law emission resulting from the hot, optically thin corona above the disc. 

In addition to both of these explanations for the soft excess it is also possible that in systems being observed at higher inclinations absorption effects could produce similarly shaped spectra (e.g. \citealt{GierlinksiDone2004,Tanaka+2004}). The existence of gas and dust clouds in the central region of AGN may act to partially obscure the view of the centre-most region. These smaller clouds of matter can therefore greatly affect the spectral shape and variability in AGN. These partial covering absorbers may mimic the signatures of relativistic reflection and soft Comptonisation via the absorption of intermediate-energy X-rays such that a soft excess is produced. 

Seyfert galaxies exhibit many types of variability over various timescales (e.g. \citealt{Grupe+2012}). Periodic behaviour, though commonly detected in X-ray binaries, is something of a rarity for AGN \citep{Zhang+2018} with only a handful of objects exhibiting evidence of periodicity in the X-ray. Some examples include RE J1034+396 \citep{Gierlinski+2008}, 1H 0707$-$495 \citep{Pan+2016}, Mrk 766 \citep{Zhang+2017}, and NGC 3516 \citep{Iwasawa+2004}. Power spectrum analyses of these sources revealed periodic variability in a narrow band simultaneous with broadband variability, with the exception of NGC 3516 which exhibited only a few periods of sinusoid-like behaviour in the light curve.

There has been much debate over the significance of the detected periods in AGN, as a number of claims of periodicity in such objects have been disproved (e.g. \citealt{Madejski+1993,BarthStern2018}). The robust measurements are often associated with quasi-periodic oscillations (QPOs) in a broad continuum band or modulations in the relativistic broadened Fe~K$\alpha$ emission line.

Markarian 530 (NGC 7603, $z=0.0295$; hereafter Mrk 530) is a bright (flux in $2-10$ keV band of $\sim2\times10^{-11}$ ergs cm$^{-2}$ s$^{-1}$) Seyfert 1.5 galaxy that has been included in numerous surveys studying various wavelength regimes (e.g. \citealt{Lal+2004,Singh+2013,Theios+2016,Malkan+2017}). Mrk 530 has a line-of-sight companion galaxy (NGC 7603B) that is at a higher redshift and connected by filaments observed in optical (e.g. \citealt{LopezGutierrez2002}). Features of possible tidal origin surrounding Mrk $530$ have been identified in optical images, indicating that the galaxy may have undergone a recent merger event \citep{Slavcheva-MihovaMihov2011}. Mrk 530 has been found to be a so-called ``changing-look'' AGN (e.g. \citealt{Matt+2003,Puccetti+2007,LaMassa+2015}) as it apparently changes from a Seyfert type 1.9 to type 1.0 \citep{Goodrich1995,Mickaelian2016}. The object has shown optical spectral variations over the course of months to years and exhibits stronger amplitudes of optical emission line variability compared to other Seyferts \citep{Kollatschny+2000}. Mrk $530$ remains relatively poorly studied with modern X-ray telescopes. This work presents an in-depth X-ray spectral analysis of Mrk 530 with \textit{Suzaku}. 

The work presented here is organised in the following manner. In Section \ref{sec:obs} we list the observations of Mrk 530 that we analysed as well as the methods used in the analysis. Section \ref{sec:timing} presents the timing variability observed and a brief look into the spectral variability. Initial spectral fits are performed in Section \ref{sec:initial-spectral-fits}. The spectral modelling for the average, soft and hard, and phase-resolved spectra are presented in Sections \ref{sec:physical-avg}, \ref{sec:spec-var}, and \ref{sec:phase-resolved}, respectively. We discuss the results more in-depth in Section \ref{sec:discussion} and finally conclude in Section \ref{sec:conclusions}. 

%%%%%%%%%%%%%%%%%%%%%%%%%%%%%%%%%%%%%%%%%%%
\section{Observations \& Data Reduction} %%
%%%%%%%%%%%%%%%%%%%%%%%%%%%%%%%%%%%%%%%%%%%
\label{sec:obs}
Mrk 530 was observed with the \textit{Suzaku} satellite starting 6 June 2012 (observation ID 707003010) in the XIS-nominal position for a total exposure time of $\sim101$ ks. Data were collected by the two front-illuminated (FI) CCDs (XIS0 and XIS3), the back-illuminated (BI) CCD (XIS1) and the HXD-PIN detectors. The  \textsc{xselect} software was used to extract spectra and light curves from the XIS detector event list files. Source spectra were extracted from a circular region centred on the source with a radius of $4'$ and background spectra were selected from a circular region off-source with a $2'$ radius. The RMF response matrices and ARF ancillary response files for the observations were generated using \textsc{xisrmfgen} and \textsc{xissimarfgen}, respectively. Spectra from XIS0 and XIS3 were merged using \textsc{addascaspec} after ensuring that the spectra were consistent. The XIS1 data were also checked for consistency though these  were not used in this analysis for the sake of simplicity. These were omitted in order to avoid inconsistencies arising from cross-calibration of the FI- and BI-CCD detectors. We favour use of the FI-CCDs due to the higher number of counts available from merging the spectra of XIS0 and XIS3.

Source and background spectra from the HXD-PIN detector were extracted using \textsc{hxdpinxbpi}. The source is detected up to $30$ keV. 

%%%%%%%%%%%%%%%%%%%%%%%%%%%%%%%%%%%%%%%%%%%
\section{Light Curves \& Hardness Ratio} %%
\begin{figure*}
    \centering
    \includegraphics[width=\textwidth]{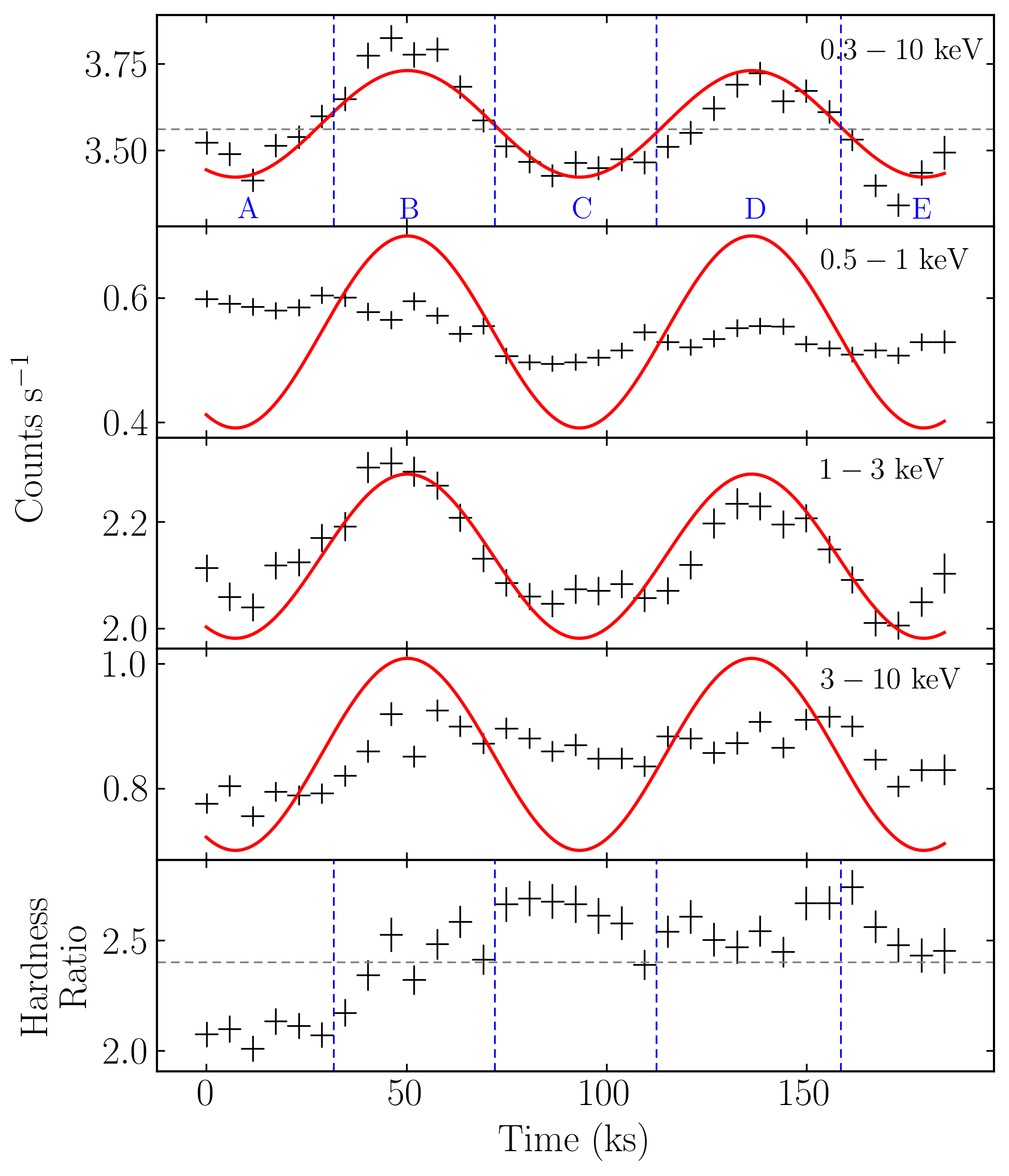}
    \caption{\textit{Top panel:} Broad-band ($0.3-10$ keV) light curve from the FI-XIS detectors fit with a sine curve. Horizontal line indicates weighted mean count rate. Vertical lines denote divisions used to create phase-resolved spectra. \textit{Middle panels:} Light curves of the $0.5-1$ keV, $1-3$ keV, and $3-10$ keV bands with the same sine function overplotted, renormalised to the average count rate of each light curve. \textit{Bottom panel:} Hardness ratio (hard/soft) between the $0.5-1$ keV (soft) and $2-10$ keV (hard) energy bands. Horizontal line indicates weighted mean hardness ratio. Vertical lines denote phase-resolved segments (A,B,C,D,E) as in top panel.}
    \label{fig:master_lc}
\end{figure*}

%%%%%%%%%%%%%%%%%%%%%%%%%%%%%%%%%%%%%%%%%%%
\label{sec:timing}
The broad-band ($0.3-10$ keV) light curve of Mrk 530 taken by the XIS0+3 detectors is shown in the top panel of Figure \ref{fig:master_lc}. The weighted average count rate is $3.561\pm0.006$ counts s$^{-1}$ as indicated by the dashed grey line in the panel. A constant fit to the light curve yields a poor fit with a statistic of $\chi^2_{\nu}= 12.65$, indicating moderate short-term variability. The fractional variability \citep{Edelson+2002} of the broad-band light curve is $3.4\pm0.5$ per cent. 

The broad-band light curve shows a curious sinusoidal behavior, displaying two clear peaks and troughs of roughly the same amplitude. However, red noise can often mimic a few cycles of sinusoidal behaviour \citep{Vaughan+2016}, and so robust detections of QPO behaviour have been those that have been detected in addition to the broadband aperiodic variability. Nonetheless, to characterise the light curve, a sine function of the form:
\begin{equation}
    \label{eq:sine}
    y = A \sin(\omega x+\phi) + c    
\end{equation}
was fit to the broad-band light curve with $A,\omega,\phi,c$ as free parameters. This fit, plotted as the red curve in the top panel of Figure \ref{fig:master_lc}, yielded a  statistic of $\chi^2_{\nu}=2.57$ and a period of $T \approx 86$ ks. 

To investigate spectral variability, light curves were made in three different energy bands. The $0.5-1$ keV, $1-3$ keV, and $3-10$ keV energy ranges were chosen to represent the soft excess, continuum emission, and hard emission, respectively. These light curves are shown in the second, third, and fourth panels of Figure \ref{fig:master_lc}. The red curve plotted over each light curve is Equation \ref{eq:sine} fit to the broad-band light curve and renormalised to the average count rate of the light curve on which it is plotted. This reveals that the seemingly periodic variability is primarily seen in the intermediate energy band between $1-3$ keV.

To further characterise the oscillatory variability of this observation, phase-resolved spectra were created by dividing the broad-band light curve into the segments denoted by vertical lines in the top panel of Figure \ref{fig:master_lc} labelled A, B, C, D, E. These segments were chosen such that each spectrum corresponded to either a peak (B, D) or trough (A, C, E) in the light curve, and the segments were relatively equally-spaced. 

The bottom panel of Figure \ref{fig:master_lc} shows the hardness ratio (HR) of the $2-10$ keV (hard) band count rate divided by the $0.5-1$ keV (soft) band count rate. The weighted average of the hardness ratio is $2.40\pm0.01$ as indicated by the dashed grey line in the panel. The weighted average fit to the hardness ratio produces a fit statistic of $\chi^2_{\nu}=10.23$ indicating significant changes in the hardness ratio over the course of the observation. Examining the shape of the hardness ratio reveals two distinct spectral states: a soft state (HR $<2.4$ and $t\lesssim50$ ks) which then transitions to a hard state (HR $>2.4$ and $t\gtrsim50$ ks). This transition is remarkably smooth and gradual over time. It should be noted that the designation of soft and hard states in this analysis should not be interpreted as major changes in the accretion geometry associated with major changes in the Eddington rate as in BH X-ray binaries; this is merely a distinction of the different intervals observed in the hardness ratio.

Time segment A of the hardness ratio is entirely in the soft state, segment B the transition state, and segments C, D, E are entirely in the hard state. These segments are divided by the blue vertical lines in the panel, which correspond to the segments labelled in the top panel of the figure.

%%%%%%%%%%%%%%%%%%%%%%%%%%%%%%%%%%
\section{Initial Spectral Fits} %%
%%%%%%%%%%%%%%%%%%%%%%%%%%%%%%%%%%
\label{sec:initial-spectral-fits}

%%%%%%%%%%%%%%%%%%%%%%%%%%%%%%%%%%
\subsection{General}
\label{sec:general}

XIS0+3 data were optimally binned as described by \cite{KastraBleeker2016} using the \textsc{python} code written by C. Ferrigno\footnote{\url{https://cms.unige.ch/isdc/ferrigno/developed-code/}}. Due to instrument calibration and sensitivity, energy channels below $0.7$ keV and above $10$ keV were omitted from the XIS0+3 data. Additionally, the energies $1.72-1.88$ keV and $2.19-2.37$ keV were omitted due to known instrument calibration issues \citep{Nowak+2011}. PIN data within the energy range of $15-30$ keV were also used in this analysis, though they were not binned due to insufficient counts. 

The X-ray spectral modelling software \textsc{xspec} v.12.9.1 \citep{Arnaud1996} was used to perform all spectral fitting. Fit quality was quantified in \textsc{xspec} using the $C$-statistic, a modification of the  Cash statistic \citep{Cash1979}, which has been been shown to be an appropriate statistic for X-ray astronomy (e.g. \citealt{Humphrey+2009,Mantz+2017}). The Galactic column density toward Mrk $530$ was kept frozen at $3.82\times10^{20}$ cm$^{-2}$ \citep{Karlberla+2005} during all spectral fitting. The cross-normalisation factor between the XIS and HXD-PIN detectors during this observation was frozen to $1.16$ as the observation of Mrk 530 was taken in the XIS nominal position\footnote{From section $5.5.11$ of Suzaku Data Reduction Guide: \url{https://heasarc.gsfc.nasa.gov/docs/suzaku/analysis/abc/node8.html}}. All errors are calculated to a $90$ per cent confidence level.

%%%%%%%%%%%%%%%%%%%%%%%%%%%%%%%%%%
\subsection{Simple Models}
\label{sec:simple-models}
\begin{figure}
    \centering
    \includegraphics[width=\columnwidth]{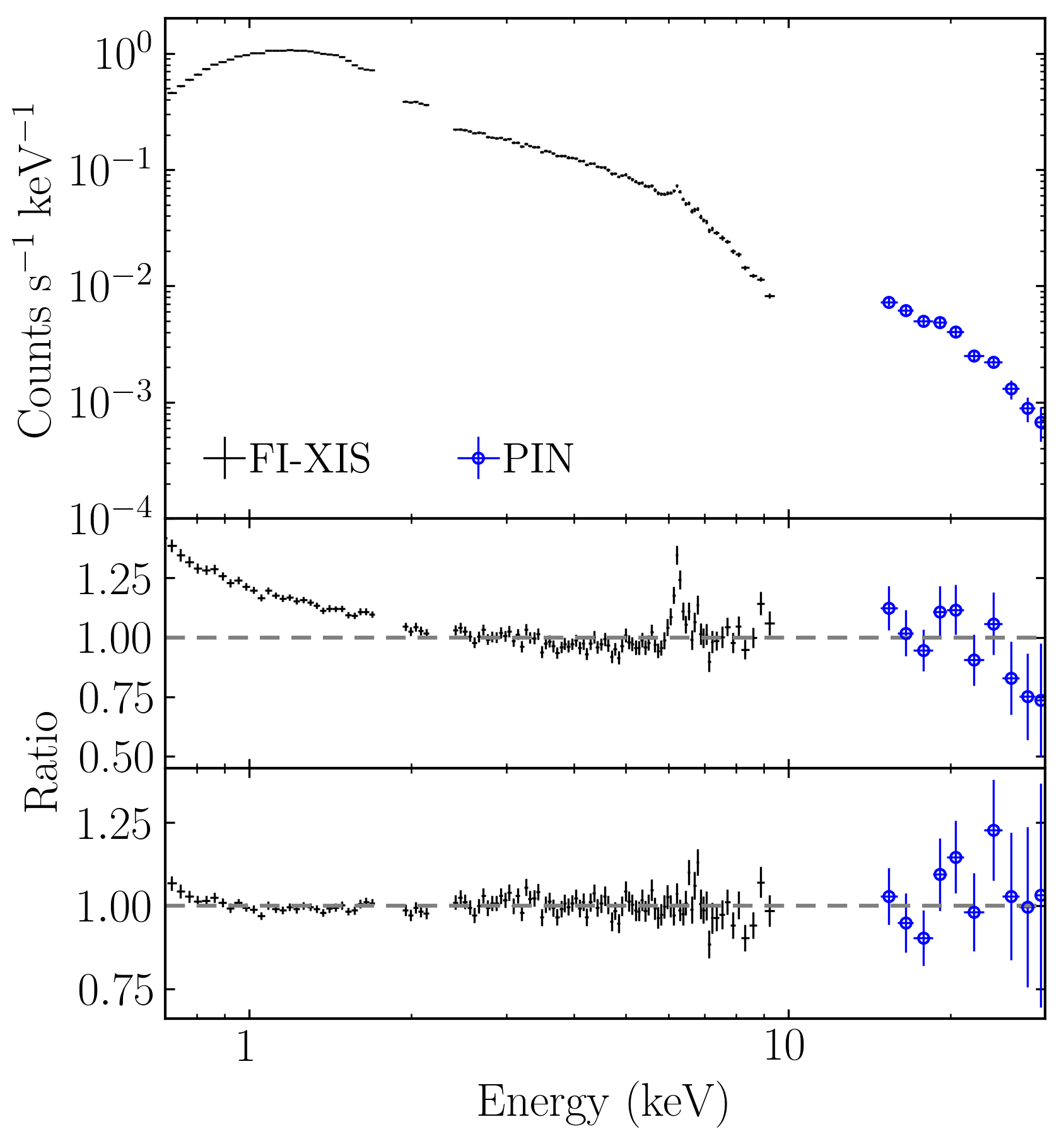}
    \caption{\textit{Top panel:} Spectrum of \textit{Suzaku} observation of Mrk 530 used in this analysis. \textit{Middle panel:} Ratio residuals of an absorbed power-law model fit from $2-10$ keV and extrapolated over $0.7-30$ keV, showing a strong soft excess below $2$ keV and a weaker hard excess between $8-15$ keV, as well as residuals in the Fe K$\alpha$ band. \textit{Bottom panel: }Ratio residuals of a blackbody model with a broad Gaussian fit to the broad-band spectrum.}
    \label{fig:extrapolated_powerlaw}
\end{figure}

The broad-band data that were used in spectral fitting are shown in the top panel of Figure \ref{fig:extrapolated_powerlaw}. To examine the spectrum, we first fit it with a power-law (model \textsc{po} in \textsc{xspec}) accounting for Galactic absorbtion along our line of sight (\textsc{tbabs}; \citealt{Wilms+2000}). This model (\textsc{tbabs*po}) was fit from $2-10$ keV to model the continuum emission, and then extrapolated to lower and higher energies to cover the full extent of the XIS and PIN detectors. The ratio residuals (data/model) of this fit are shown in the middle panel of Figure \ref{fig:extrapolated_powerlaw}. This reveals a strong soft excess present in Mrk 530 below $2$ keV, as well as a weak excess between $8-15$ keV and excess emission in the Fe K$\alpha$ band.

To model this soft excess we first used the simplest model, in which the soft excess is described by thermal blackbody radiation from the accretion disc. We included in this model a single redshifted Gaussian emission line that was left free to vary in both energy and width. The model (\textsc{const*tbabs*(po+bb+zgauss)} in \textsc{xspec}) was fit to the broadband data, yielding the ratio residuals shown in the bottom panel of Figure \ref{fig:extrapolated_powerlaw}. This fit, with a statistic of $C/\mathrm{d.o.f.}=189.60/166$, produced a broadened emission line at $\sim6.4$ keV with a width of $\sigma \sim85$ eV. The fit yielded a blackbody temperature of $380\pm30$ eV, which rules out this model as physically realistic, as the temperature is too high to originate from the accretion disc.

To characterise the data and the nature of the soft excess, we explored three different physical models: a blurred reflection model (Section \ref{sec:blur}), a partial covering model (Section \ref{sec:abs}), and a soft Comptonisation model (Section \ref{sec:optx}). 

%%%%%%%%%%%%%%%%%%%%%%%%%%%%%%%%%%
\subsection{The Fe K$\alpha$ band}
\label{sec:feka}
\begin{figure}
    \centering
    \includegraphics[width=\columnwidth]{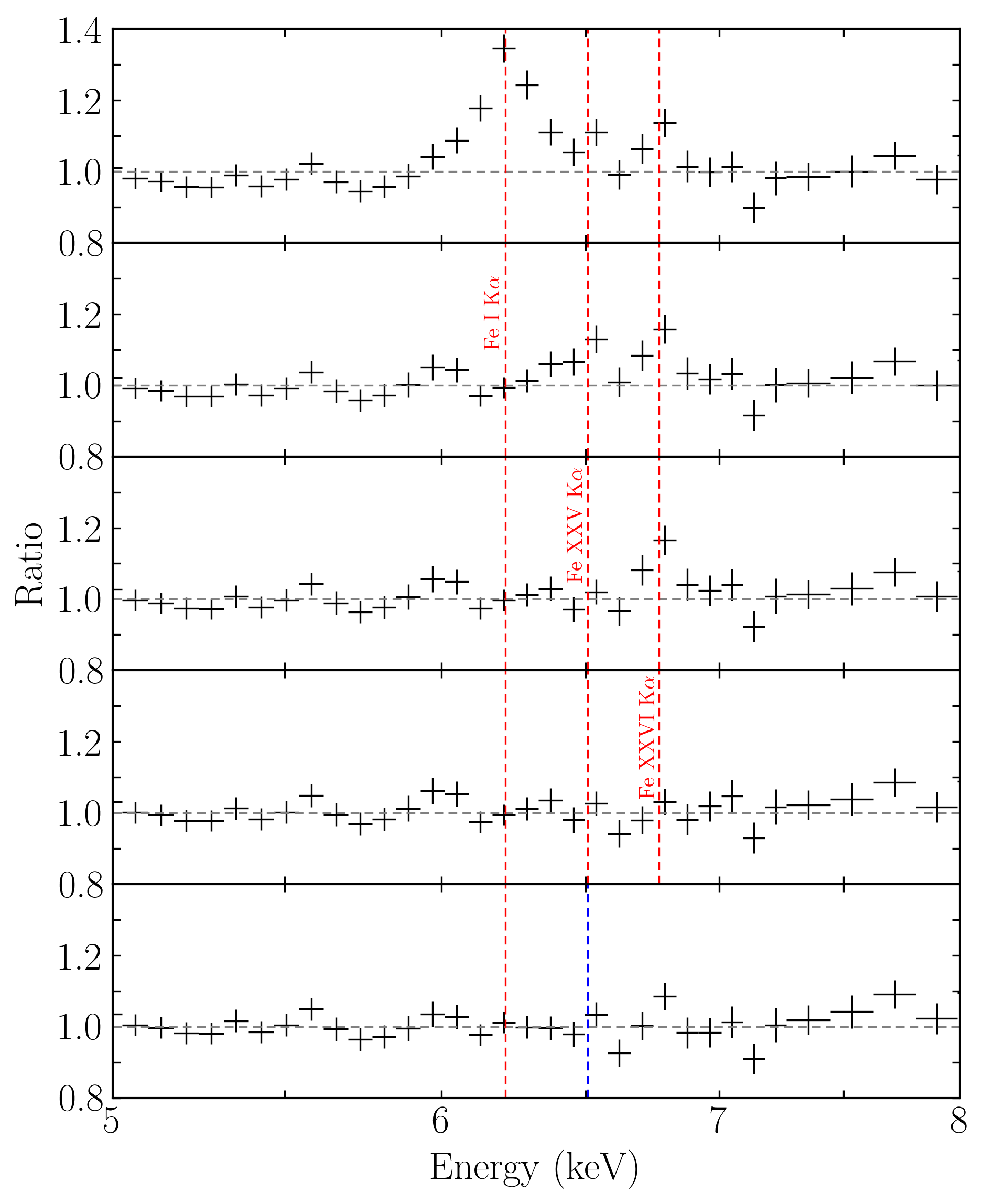}
    \caption{\textit{Top panel:} Ratio residuals of the $2-10$ keV region fit with absorbed power-law. \textit{Second panel:} As above, but model includes a narrow Gaussian line fixed at $6.4$ keV to fit the Fe I K$\alpha$ emission line. \textit{Third panel:} As above, with another Gaussian line at $6.7$ keV to fit the Fe XXV K$\alpha$ emission line. \textit{Fourth panel:} As above, with a narrow Gaussian line at $6.97$ keV to fit the Fe XXVI K$\alpha$ emission line. Bottom panel: Ratio residuals of the $2-10$ keV region fit absorbed power-law with narrow Gaussian at $6.4$ keV and a broad Gaussian at $6.7$ keV. Vertical dashed lines indicate observed energies to which these lines were fit.}
    \label{fig:fe_lines}
\end{figure}

Fitting the $2-10$ keV spectrum with only an absorbed power-law leaves strong residuals in the Fe K$\alpha$ region (around $6-7$ keV) (Figure \ref{fig:fe_lines}, top panel). Though fit from $2-10$ keV, this figure displays only the $5-8$ keV band to clearly see changes in the Fe K$\alpha$ region. This fit exhibits strong residuals in this region and yields a poor fit with a statistic of $C/\mathrm{d.o.f.}=327.98/101$.

As shown in the bottom panel of Figure \ref{fig:extrapolated_powerlaw}, a single broad Gaussian is unable to account for the residuals in the Fe K$\alpha$ region. We thus consider a multiple narrow-line scenario. First, a narrow emission line was added to the absorbed power-law model (\textsc{tbabs*(po+zgauss)}) at an intrinsic line energy of $6.4$ keV. This line, corresponding to the K$\alpha$ emission lines from neutral (or low ionized) iron (Fe I K$\alpha$), was fixed to a width of $\sigma=1$ eV, as were all narrow lines. This Gaussian component improved the fit, yielding $C/\mathrm{d.o.f.}=170.40/100$. The ratio residuals of this fit are shown in the second panel of Figure \ref{fig:fe_lines}. The excess residuals exist blueward of $6.4$ keV, which may not be fully consistent with relativistic blurring. To further improve the residuals, a second narrow Gaussian feature was fit to an intrinsic energy of $6.7$ keV. The ratio residuals of this fit are shown in the third panel of Figure \ref{fig:fe_lines}. This line corresponds to a K$\alpha$ emission line from He-like iron (Fe XXV K$\alpha$). This additional component betters the fit, yielding $C/\mathrm{d.o.f.}=152.66/99$. A narrow Gaussian line was then fit to an intrinsic energy of $6.97$ keV, which corresponds to a K$\alpha$ emission line from H-like iron (Fe XXVI K$\alpha$). This fit, the ratio residuals of which are shown in the fourth panel of Figure \ref{fig:fe_lines}, further improved the fit, yielding a statistic of $C/\mathrm{d.o.f.}=132.26/98$. Vertical lines in Figure \ref{fig:fe_lines} denote the observed energies at which these Gaussian profiles were fit. We modelled the Compton shoulder with a Gaussian at an intrinsic energy of $6.3$ keV and fixed to a normalisation one tenth that of the Fe I K$\alpha$ line \citep{Matt+1991}. However, while the data were consistent with this, this addition did not significantly improve the fit ($C$/d.o.f $=131.69/98$). Additionally, it should be noted that a Gaussian at 6.3 keV is a crude approximation of the Compton shoulder.

The data can support an alternative interpretation to the single broad emission line and the multiple narrow line scenarios described above. Fitting a narrow line fixed at $6.4$ keV with a width of $\sigma=1$ eV and fitting a second Gaussian with energy and width left free to vary produces a broad line at $6.7$ keV with a width of $\sigma=400\pm200$ eV. This fit, with a statistic of $C/\mathrm{d.o.f.}=134.24/97$, can also reasonably account for the Fe K$\alpha$ excess. The ratio residuals of this fit are shown in the bottom panel of Figure \ref{fig:fe_lines}.

%%%%%%%%%%%%%%%%%%%%%%%%%%%%
\section{Physical Models} %%
%%%%%%%%%%%%%%%%%%%%%%%%%%%%
\label{sec:physical-avg}

%%%%%%%%%%%%%%%%%%%%%%%%%%%%%%%%%%%
\subsection{Blurred reflection}
\label{sec:blur}

The curvature present in the average broadband spectrum of Mrk 530 may be interpreted in a number of ways. One scenario that can account for the shape is relativistic reflection off of the inner region of the accretion disc. We therefore begin by fitting the average broadband X-ray spectrum with blurred ionised reflection (\textsc{relxill}, \citealt{Garcia+2014,Dauser+2014}).

The fit parameters are shown in Table \ref{tab:avg_master}. The model consists of a cut-off power law, blurred ionised reflector, and neutral reflector; in \textsc{xspec} the model is: \textsc{const*tbabs*(cflux*cutoffpl+const*cflux*relxill+pexmon} \textsc{+zgauss+zgauss)}. The cut-off energy of the power-law component was frozen to $E_{\mathrm{cut}} = 300$ keV as the fit was unable to constrain it when allowed to vary. We have assumed the black hole spin parameter is maximal, therefore allowing the inner radius of the accretion disc to extend to the minimum allowed by the model. The disc is assumed to be illuminated such that emissivity index is constant at $q=3$ over the entire disc. Allowing these parameters to vary freely found them to be poorly constrained, thus they are frozen at the listed values. Iron abundance in the accretion disc was found to be consistent with solar and therefore it is frozen at $A_{\mathrm{Fe}} = 1.0$. The use of \textsc{cflux} allows for the flux of the power law and reflection components to be set equal, letting the modulation of the flux from the blurred reflection component to be computed external to the model with a simple constant component. 

Reflection from the cold distant torus was modelled using \textsc{pexmon} \citep{Nandra+2007}. This model includes neutral Compton reflection and self-consistent Fe and Ni emission lines, including the $6.4$ keV Fe K$\alpha$ emission line and the Compton shoulder. Two additional Gaussian components were added to this model to account for the ionised iron emission lines at $6.7$ keV and $6.97$ keV. The model components are shown in the top left panel of Figure \ref{fig:avg_eem_ra}.

The ratio residuals of the blurred reflection model fit to the average spectrum are shown in the bottom left panel of Figure \ref{fig:avg_eem_ra}. The fit is of good statistical quality, with $C/\mathrm{d.o.f.} = 171.18/166$. 
This model describes a scenario in which a moderately steep power law illuminates the accretion disc and torus. The accretion disc emission is highly ionised and is a significantly weaker component than the power-law, as the reflection fraction is low ($R=0.22$). The inclusion of the cold distant reflection component is necessary to model the high energy curvature in the spectrum. However, even with the inclusion of both reflection components the ratio residuals above $\sim8$ keV deviate significantly from the model, indicating further curvature in the spectrum.

%%%%%%%%%%%%%%%%%%%%%%%%%%%%%%%%
\subsection{Partial covering}
\label{sec:abs}

An alternate explanation to the curvature observed in the average spectrum of Mrk 530 is partial covering of the inner region by ionised and/or neutral clouds obscuring the system. The partial covering absorption was modelled by using the \textsc{xspec} model \textsc{zxipcf}. The fit parameters for such a scenario are given in Table \ref{tab:avg_master}. The model consists of two partial covering absorbers convolved with a cut-off power-law as well as a neutral reflection component; the \textsc{xspec} model is: \textsc{const*tbabs*(zxipcf*zxipcf*cflux*cutoffpl+pexmon+} \textsc{zgauss+zgauss}). During the fitting procedure it was found that only one absorber required ionisation to be left free while the other was consistent with neutral material. The neutral material was modelled by freezing the ionisation parameter ($\xi=4\pi F/n$, where $F$ is the flux of the illuminating radiation and $n$ the gas density) to $\log\xi = -3$, the minimum allowed by the model. Here \textsc{cflux} is used to give the unabsorbed flux of the power-law. The Gaussian components are again added to fit the $6.7$ keV and $6.97$ keV emission features. The model components are shown in the top centre panel of Figure \ref{fig:avg_eem_ra}. 

The data are over-fit by this absorption model, with $C/\mathrm{d.o.f.} = 143.82/163$. The ratio residuals, shown in the bottom centre panel of Figure \ref{fig:avg_eem_ra}, reveal that the high-energy curvature is minimised with this model, though at the lowest energies ($\lesssim0.9$ keV) deviations from the model can be seen. This is likely indicative of a stronger soft excess than is predicted by the model. The lack of sensitivity below $0.7$ keV in \textit{Suzaku} makes it difficult to investigate further.

In this scenario a neutral, high column-density cloud partially obscures a large amount of the power-law emission while a lightly ionised, lower column-density cloud partially obscures less of the emission. The power-law itself is much steeper than in the blurred reflection case and the cold reflection component is also brighter by a factor of two. 

%%%%%%%%%%%%%%%%%%%%%%%%%%%%%%%%%%%
\subsection{Soft Comptonisation}
\label{sec:optx}
\begin{figure*}
    \centering
    \includegraphics[width=\textwidth]{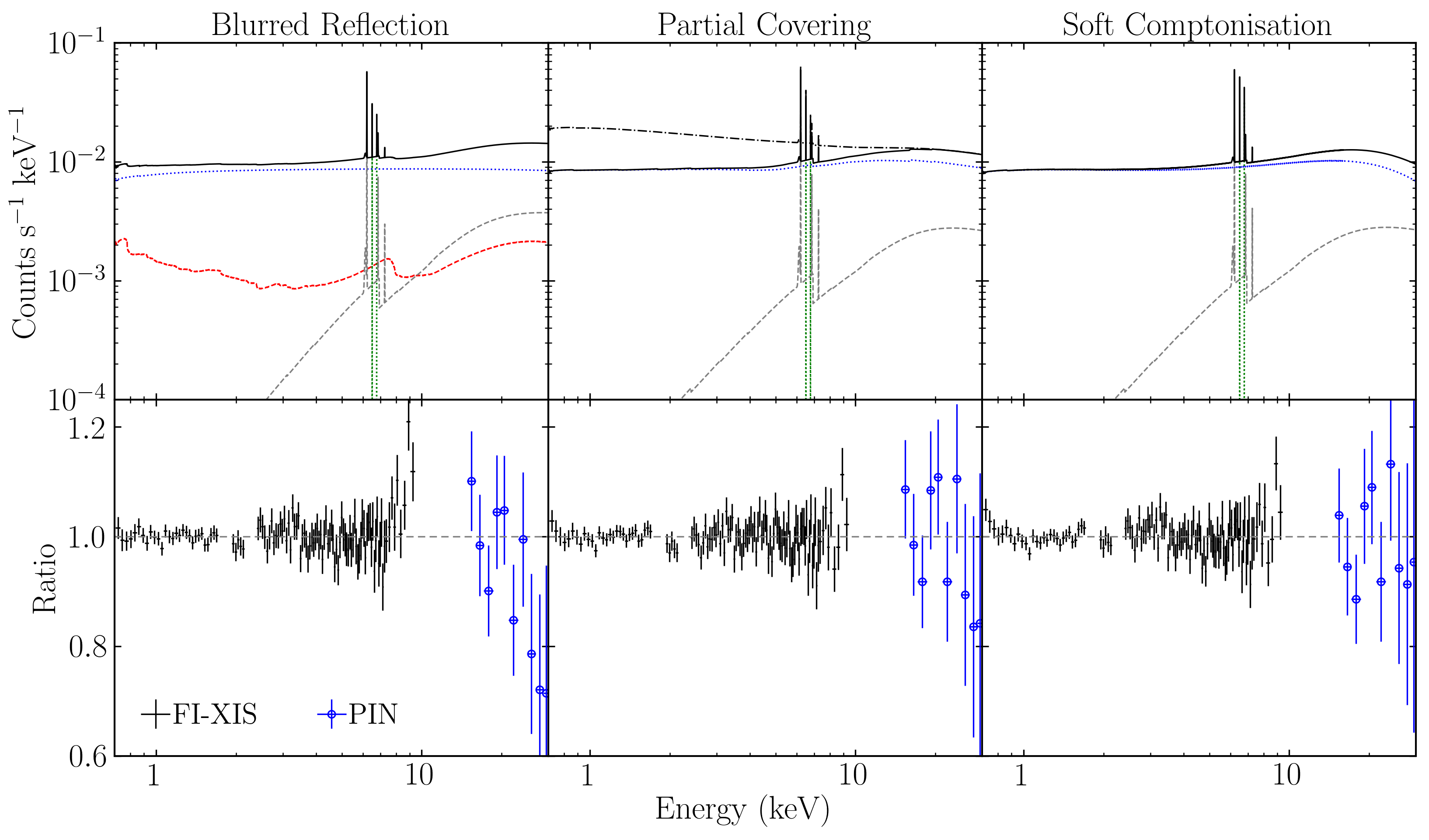}
    \caption{Models (top row) and ratio residuals (bottom row) of the three tested models: blurred reflection (left), partial covering (middle), and soft Comptonisation (right). Blurred reflection model components include power-law (blue), ionised iron lines (green), reflection from torus (grey), and reflection from accretion disc (red). Parital covering model components include absorbed power-law (blue), ionised iron lines (green), and reflection from torus (grey). Unabsorbed spectrum is shown by the black dot-dash curve. Soft Comptonisation model includes soft Comptonised power-law (blue), ionised iron lines (green), and reflection from torus (grey).}
    \label{fig:avg_eem_ra}
\end{figure*}

The curvature in the spectrum of Mrk $530$ can also be described by a soft Comptonisation scenario (e.g. \citealt{Matt+2014}), which was modelled with \textsc{optxagnf} \citep{Done+2012}. This model describes the soft excess with a thermal Comptonisation disc component that is optically thick with a low temperature, and models the continuum emission above $2$ keV with an optically thick, high temperature thermal Comptonisation. The torus was again modelled with \textsc{pexmon} and the two ionised iron emission lines were included. The broad-band continuum was modelled by \textsc{const*tbabs*(optxagnf+pexmon+zgauss+zgauss)}, the fit parameters of which are shown in Table \ref{tab:avg_master}. Figure \ref{fig:avg_eem_ra} displays the model components (upper right panel) and fit ratio residuals (bottom right panel).

During the fit, the following five \textsc{optxagnf} parameters were left free to vary:  the Eddington ratio ($L/L_{\mathrm{edd}}$), electron temperature in the soft component ($kT_e$), optical depth of soft component ($\tau$), photon index of hard component ($\Gamma$), and fraction of the power in the hard component ($f_{\mathrm{pl}}$). Black hole spin was frozen to the maximum value as a range of values for spin were implemented and that was found to produce the best fit. The coronal radius $r_{cor}$ was fixed to $10$ $r_g$ as the Eddington ratio was found to be insensitive to radii less than that. The Eddington ratio was found to become lower with increasing  $r_{\mathrm{cor}}$ until about $10$ $r_g$, at which point the Eddington ratio became insensitive to the parameter. Since \cite{Singh+2011} found an Eddington ratio for Mrk $530$ of  $\log(L/L_{\mathrm{edd}})=-2.06$ which was much lower than this model would allow, we thus froze $r_{\mathrm{cor}}$ to the value which minimized the Eddington ratio. The radius of the outer disc was frozen to $1000$ $r_g$ as the fit was insensitive to this parameter. The cutoff energy of the \textsc{pexmon} model was fixed to $300$ keV for consistency between the three compared models.  

This soft Comptonisation model over-fits the data with $C/\mathrm{d.o.f}=159.81/165$. The same trend is observed in these residuals as those from the partial covering model: the high-energy curvature is minimised, but there remains a soft excess at the lowest observable energies.

\begin{table*}
    \centering
    \caption{Parameter values for the three models fit to the average spectrum (FI-XIS \& PIN) for Mrk 530. All values without uncertainties are frozen during the fitting, as allowing them to be free did not improve the fit significantly.         \textsc{zgauss} norm in units of source frame photons cm$^{-2}$ s$^{-1}$.
        \textsc{pexmon} norm in units of observed frame photons keV$^{-1}$ cm$^{-2}$ s$^{-1}$ at 1 keV of the cutoff power-law. \textsc{cflux} norm in units of source frame photons keV$^{-1}$ cm$^{-2}$ s$^{-1}$ at 1 keV.}
    \begin{tabular}{ccccc}
        \hline
        Model           & Parameter                 & Blurred  Reflection  & Partial Covering  & Soft Comptonisation  \\ 
        \hline
        \hline
        \textsc{cflux}  & $E_{\mathrm{min}}$ [keV] & $0.1$ &$0.1$ &$-$ \\
                        & $E_{\mathrm{max}}$ [keV] & $100$ &$100$ &$-$ \\
                        & $\log F$ [erg cm$^{-2}$ s$^{-1}$] & $-10.08^{+0.02}_{-0.03}$ &$-9.72\pm0.06$ &$-$ \\
        \hline
        \textsc{cutoffpl}   & $\Gamma$                  & $1.97\pm0.02$ & $2.22\pm0.06$  & $-$  \\
                            & $E_{\mathrm{cut}}$ [keV]  &$300$ & $300$    & $-$ \\
                            & norm ($\times10^{-2}$)    & $1$ & $1$  & $-$ \\
        \hline 
        \textsc{const}      & $R$                       &$0.22^{+0.08}_{-0.07}$ & $-$ & $-$ \\
        \hline 
        \textsc{relxill}    & $q_{\mathrm{in}}$         & $3.0$     & $-$  & $-$ \\
                            & $q_{\mathrm{out}}$        & $3.0$     & $-$  & $-$ \\
                            &  $R_{\mathrm{break}}$ $r_g$     & $15.0$     & $-$  & $-$ \\
                            & $a$                       & $0.998$    & $-$  & $-$ \\
                            & $i$ [$^{\circ}$]          & $61.5$ $^{a}$     & $-$  & $-$ \\
                            & $R_{\mathrm{in}}$ [$r_g$]      & $1.235$     & $-$  & $-$ \\
                            & $R_{\mathrm{out}}$ [$r_g$]      & $400$     & $-$  & $-$ \\
                            & $\Gamma$                  & $1.97$ $^*$     & $-$  & $-$ \\
                            & $\log \xi$ [erg cm s$^{-1}$]   & $2.8\pm0.2$ & $-$ & $-$ \\
                            & $A_{\mathrm{Fe}}$                  & $1.0$             & $-$  & $-$ \\
                            & $E_{\mathrm{cut}}$ [keV]        & $300$  & $-$  & $-$ \\
                            & $R_{\mathrm{rel}}$               & $-1$ & $-$  & $-$ \\
                            & norm                      & $1$ & $-$  & $-$ \\
        \hline
        \textsc{zxipcf} & N$_H$ ($\times10^{22}$) [cm$^{-2}$]   & $-$ & $22^{+7}_{-3}$   & $-$ \\
                        & $\log\xi$ [erg cm s$^{-1}$]           & $-$ & $-3$ & $-$ \\
                        & CF                                    & $-$ & $0.41\pm0.05$  & $-$\\ 
        \hline
        \textsc{zxipcf} & N$_H$ ($\times10^{22}$) [cm$^{-2}$]   & $-$ & $1.9^{+0.8}_{-0.4}$   & $-$ \\
                        & $\log\xi$ [erg cm s$^{-1}$]           & $-$ & $-0.6^{+1.0}_{-0.4}$ & $-$ \\
                        & CF                                    & $-$ & $0.26^{+0.07}_{-0.05}$  & $-$\\ 
        \hline
        \textsc{optxagnf}   &$M_{\mathrm{BH}}$ [$M_{\odot}$] &$-$&$-$& $1.15\times10^8$ $^{b}$  \\
			                & $D$ [Mpc] &$-$&$-$& $118.3$ $^{c}$   \\
			                & $\log L/L_{\mathrm{Edd}}$ &$-$&$-$& $ -1.74 ^{+0.03}_{-0.02}$   \\
			                & $a$ &$-$&$-$& $0.998$\\
							& $r_{\mathrm{cor}}$ [$r_g$] &$-$&$-$& $10$ \\
							& $r_{\mathrm{out}}$ [$r_g$] &$-$&$-$& $1000$ \\
							& $kT_{\mathrm{e}}$ [keV] &$-$&$-$& $5.5\pm1$ \\
							& $\tau$ &$-$&$-$& $5.75^{+4}_{-2}$ \\
							& $\Gamma$ &$-$&$-$& $2.22 ^{+0.11}_{-0.08}$ $^*$  \\
							& $f_{\mathrm{pl}}$ &$-$&$-$& $ 0.83^{+0.09}_{-0.11}$\\
							& $k$ &$-$&$-$& $1$ \\                            
        \hline
        \textsc{pexmon}     & $\Gamma$                  & $1.97^*$ &  $2.22^*$      &   $2.22^*$    \\
                            & $E_{\mathrm{cut}}$ [keV]  & $300$  & $300$      & $300$   \\
                            & $R_{\mathrm{rel}}$                  & $-1$ & $-1$      &  $-1$    \\
                            & $A$                       & $1.0$ & $1.0$    & $1.0$ \\
                            & $A_{\mathrm{Fe}}$         &$1.0$ & $1.0$    & $1.0$ \\
                            & $i$ [$^{\circ}$]          & $61.5$ $^a$ & $61.5$ $^a$     & $61.5$ $^a$       \\
                            & norm ($\times10^{-3}$)    & $5.5\pm0.9$ & $10\pm2$ &  $10 ^{+4}_{-2}$ \\
        \hline
        \textsc{zgauss}     & $E$ [keV]                 & $6.7$ &$6.7$ &$6.7$  \\
                            & $\sigma$ [eV]            & $1$ & $1$ & $1$ \\
                            & norm ($\times10^{-6}$)    & $4\pm3$ &$<4.9_p$ &$4\pm3$ \\
                            & EW [eV]                       & $14^{+11}_{-9}$ & $8^{+10}_{-8}$ &$14^{+10}_{-9}$ \\
        \hline
        \textsc{zgauss}     &    $E$ [keV]                 & $6.97$ &$6.97$ &$6.97$  \\
                            & $\sigma$ [eV]            & $1$ & $1$ & $1$ \\
                            & norm ($\times10^{-6}$)   & $3\pm3$ &$<3.7_p$ & $3\pm3$ \\
                            & EW [eV]                       & $11\pm10$ & $4^{+13}_{-4}$ & $12^{+10}_{-11}$ \\
        \hline
        \hline   
        Fit Quality & $C/\mathrm{d.o.f.}$   &  $171.18/166$ & $143.82/163$ &  $159.81/165$      \\
                    &                       &  $1.03$       & $0.88$        &  $0.97$     \\
        \hline
        \multicolumn{5}{c}{\textit{Note: } Values with $^*$ are tied to power-law (\textsc{cutoffpl}). $p$ denotes lower limit is pegged at zero.} \\
        \multicolumn{5}{c}{\textit{References:} $(a)$ \citealt{Monje+2011} $(b)$ \citealt{Woo+2012} $(c)$ \citealt{Lal+2011}}

    \end{tabular}
    \label{tab:avg_master}
\end{table*}

%%%%%%%%%%%%%%%%%%%%%%%%%%%%%%%%%%%%%%%%%%%%%%%%
\section{Soft to hard state spectral changes} %%
%%%%%%%%%%%%%%%%%%%%%%%%%%%%%%%%%%%%%%%%%%%%%%%%
\label{sec:spec-var}
As indicated by the hardness ratio in the bottom panel of Figure \ref{fig:master_lc}, there are two distinct spectral states present in this observation. Spectra were made of these two states, henceforth called the soft spectrum (which covers the same time range as segment A) and hard spectrum (which covers the time intervals C to E). Spectral fitting of the above three models was performed on these spectra in an effort to understand the driver of the variability between these spectral states. Section \ref{sec:reflection_hardsoft} contains the blurred reflection fits to the soft and hard states, Section \ref{sec:pc_hardsoft} the partial covering fits, and Section \ref{sec:compt_hardsoft} the soft Comptonisation fits. 

\begin{figure}
    \centering
    \includegraphics[width=\columnwidth]{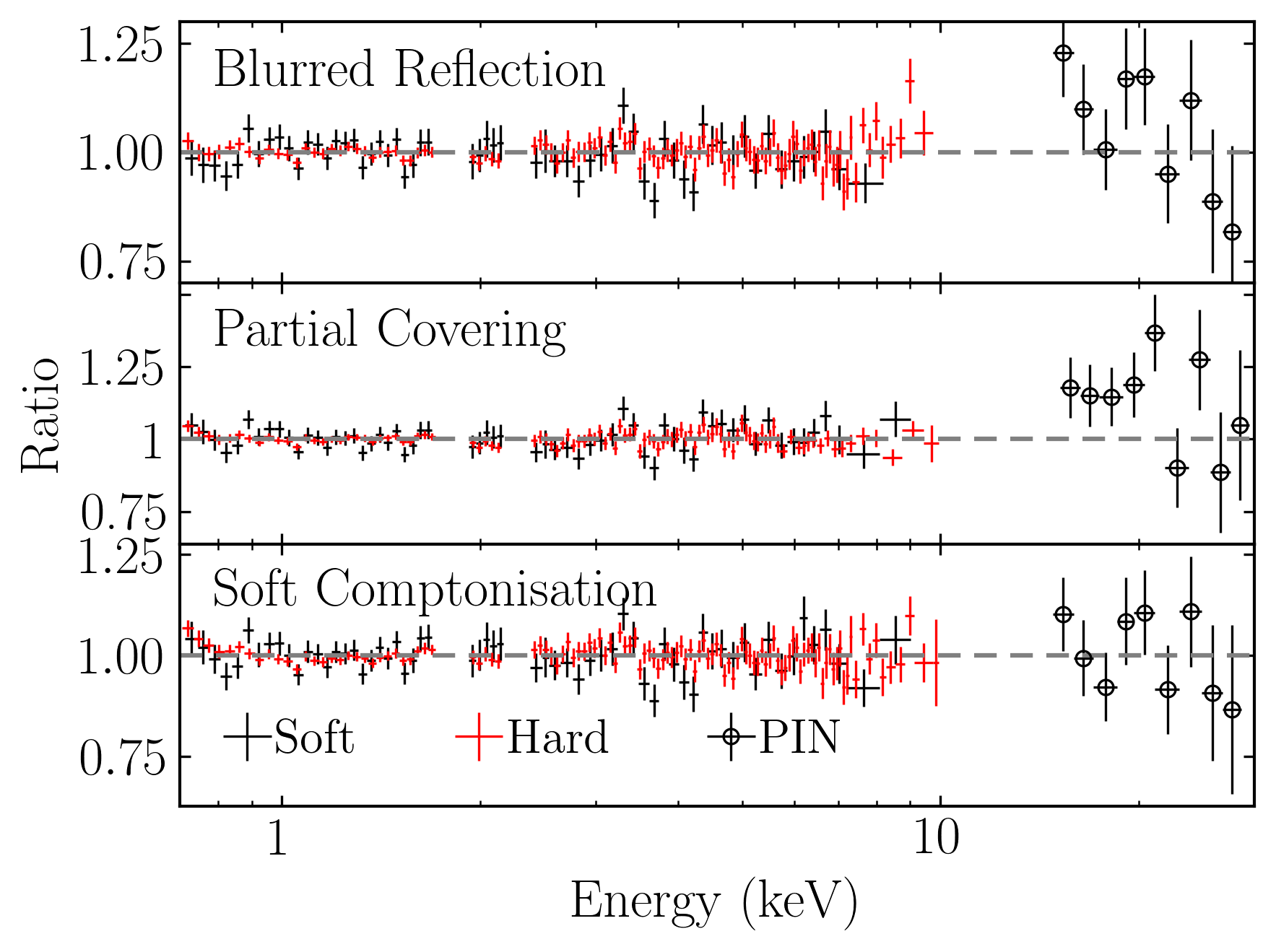}
    \caption{Ratio residuals of the soft and hard spectral states fit with the three investigated models.}
    \label{fig:softhard}
\end{figure}

\begin{table*}
    \centering
    \caption{Parameters obtained by fitting the soft and hard spectra with the three described models. A value in the middle of the column denotes a parameter that was linked between the two states.}
    \begin{tabular}{cccccccc}
        \hline
                        &               & \multicolumn{2}{c}{Blurred Reflection}    & \multicolumn{2}{c}{Partial Covering }     & \multicolumn{2}{c}{Soft Comptonisation } \\
        Model           & Parameter     & Soft  & Hard                              & Soft  & Hard                              & Soft  & Hard    \\ 
        \hline
        \hline
        \textsc{cflux}    & $\log F_{0.1-100}$ [erg cm$^{-2}$ s$^{-1}$] & \multicolumn{2}{c}{$-10.12^{+0.01}_{-0.02}$} & \multicolumn{2}{c}{$-9.74\pm0.02$} & \multicolumn{2}{c}{$-$} \\
        \hline
        \textsc{cutoffpl} & $\Gamma$ & $2.03\pm0.02$ & $1.99^{+0.01}_{-0.02}$ & \multicolumn{2}{c}{$2.27\pm0.03$} & \multicolumn{2}{c}{$-$}\\
        \hline
        \textsc{const}    & $R$ & $0.34^{+0.05}_{-0.04}$ & $0.22^{+0.06}_{-0.05}$ & \multicolumn{2}{c}{$-$} &\multicolumn{2}{c}{$-$}\\
        \hline
        \textsc{relxill}  & $\log\xi$ [erg cm s$^{-1}$]   & \multicolumn{2}{c}{$2.7\pm0.1$} & \multicolumn{2}{c}{$-$} & \multicolumn{2}{c}{$-$}\\
        \hline
        \textsc{zxipcf}   & $N_{\mathrm{H}}$ ($\times10^{22}$) [cm$^{-2}$]   & \multicolumn{2}{c}{$-$}   &                                        \multicolumn{2}{c}{$22^{f}$} & \multicolumn{2}{c}{$-$}  \\
                          & $\log\xi$  [erg cm s$^{-1}$]     & \multicolumn{2}{c}{$-$} & \multicolumn{2}{c}{$-3^{f}$} & \multicolumn{2}{c}{$-$} \\
                          & CF                        & \multicolumn{2}{c}{$-$} & \multicolumn{2}{c}{$0.41^{f}$} & \multicolumn{2}{c}{$-$}\\      
        \hline
        \textsc{zxipcf}   & $N_{\mathrm{H}}$ ($\times10^{22}$) [cm$^{-2}$]   & \multicolumn{2}{c}{$-$} &                            \multicolumn{2}{c}{$1.9^{f}$} & \multicolumn{2}{c}{$-$}  \\
                          & $\log\xi$ [erg cm s$^{-1}$]      & \multicolumn{2}{c}{$-$} & \multicolumn{2}{c}{$-0.6^{f}$} &\multicolumn{2}{c}{$-$} \\
                          & CF                        & \multicolumn{2}{c}{$-$} & $0.20\pm0.04$ & $0.29\pm0.03$ & \multicolumn{2}{c}{$-$}\\      
        \hline
        \textsc{optxagnf} & $\log L/L_{\mathrm{Edd}}$ &\multicolumn{2}{c}{$-$} &\multicolumn{2}{c}{$-$}&  $-1.62\pm0.04$ &$-1.74\pm0.01$\\ 
                          & $kT_{\mathrm{e}}$ [keV] &\multicolumn{2}{c}{$-$}&\multicolumn{2}{c}{$-$}&\multicolumn{2}{c}{$7^{+2}_{-1}$} \\
						  & $\tau$ &\multicolumn{2}{c}{$-$}&\multicolumn{2}{c}{$-$} &\multicolumn{2}{c}{$4.8^{+0.8}_{-0.7}$}\\
						  & $\Gamma$ &\multicolumn{2}{c}{$-$}&\multicolumn{2}{c}{$-$} & $2.43^{+0.07p}_{-0.08}$ & $2.27^{+0.03}_{-0.04}$\\
					      & $f_{\mathrm{pl}}$ &\multicolumn{2}{c}{$-$}&\multicolumn{2}{c}{$-$} &\multicolumn{2}{c}{$0.81\pm0.03$}\\
        \hline
        \hline
        Fit Quality & $C/\mathrm{d.o.f.}$  & \multicolumn{2}{c}{$292.39/285$}    & \multicolumn{2}{c}{$298.16/287$}  & \multicolumn{2}{c}{$299.69/284$}  \\
                    &                      & \multicolumn{2}{c}{$1.03$}          & \multicolumn{2}{c}{$1.04$}        & \multicolumn{2}{c}{$1.06$}  \\        
        \hline
        \multicolumn{8}{c}{\textit{Note: } $f$ denotes parameter is fixed, $p$ denotes parameter error is pegged at upper limit.} \\
    \end{tabular}
    \label{tab:softhard}
\end{table*}

%%%%%%%%%%%%%%%%%%%%%%%%%%%%%%%%%%
%\subsubsection{Blurred reflection}
\subsection{Blurred reflection}
\label{sec:reflection_hardsoft}
The soft and hard spectra were fit with the same blurred reflection model as in Section \ref{sec:blur}, keeping all parameters frozen to the value obtained from fitting the average spectrum, with the exception of those listed in Table \ref{tab:softhard}. The table shows the parameter values for those that were left free to vary between the soft and hard states, as well as the free parameters that were linked between the two states. 

Notably, to account for the spectral change between soft and hard we require the photon index and reflection fraction free to vary between the states. This produces a good statistical fit of $C/\mathrm{d.o.f.} = 292.39/285$ and the ratio residuals shown in the top panel of Figure \ref{fig:softhard}. By simultaneously hardening the power law and decreasing the reflection fraction (i.e. reducing the soft band curvature) we are able to harden the overall spectrum, therefore explaining the spectral state change. 

We tested for other parameter combinations to explain the spectral state change. For example, varying the photon index and ionisation parameter found that the change could be accounted for, though the fit is worse than the presented one. It was found, however, that by allowing ionisation parameter and reflection fraction free to vary between the spectra and linking the photon index in the soft and hard state, the spectral change could not be easily accounted for. That is to say that changes solely in the reflection component cannot cause the observed state change and therefore the photon index of the power law must change during the transition from the soft state to hard state.

%%%%%%%%%%%%%%%%%%%%%%%%%%%%%%%%
%\subsubsection{Partial covering}
\subsection{Partial covering}
\label{sec:pc_hardsoft}
The soft and hard spectra were then fit with the partial covering model (Section \ref{sec:abs}), again keeping all parameters frozen to the value obtained from fitting the average spectrum with the exception of those listed in Table \ref{tab:softhard} (where frozen parameters are also indicated with a superscript $f$). Here we find that by allowing covering fraction of the ionised absorber to vary between the two states we can model the spectral state change quite well ($C/\mathrm{d.o.f.} = 298.16/287$) as shown by the ratio residuals in the middle panel of Figure \ref{fig:softhard}. By increasing the covering fraction between the soft and hard states the lower energy ranges of the spectrum become more absorbed, thus hardening the spectrum and producing the state change. 

Here we tested a wide variety of parameter combinations in order to explain the spectral variability. At no point was the column density of either absorber found to change significantly between states, hence the freezing of these values in the final fit. The same was found to be true for the ionisation parameters of both absorbers as well as the covering fraction of the neutral absorber. Allowing the photon index of the power-law to vary between states was not found to improve the fit significantly. Thus, the simplest explanation of the spectral state change using a partial covering model describes a scenario in which the ionised absorber obscures more of the primary X-ray emission therefore hardening the spectrum. 

%%%%%%%%%%%%%%%%%%%%%%%%%%%%%%%%%%%
%\subsubsection{Soft Comptonisation}
\subsection{Soft Comptonisation}
\label{sec:compt_hardsoft}
Finally, the soft Comptonisation model was fit to the soft and hard spectra, the results of which are shown in Table \ref{tab:softhard}. The best fit was obtained by letting photon index and Eddington ratio free to vary between the soft and hard states. Freeing only one of those parameters (e.g. \citealt{Mallick+2017}) resulted in poor fits. Electron temperature, optical depth, $f_{\mathrm{pl}}$, as well as normalisation of the torus were free parameters which were linked between the two states. All other parameter values were kept frozen to the values obtained from the fitting the average. Ratio residuals of this fit are shown in the bottom panel of Figure \ref{fig:softhard}.

We see that changes in the Eddington ratio and power-law index are able to model the spectral changes reasonably well with a statistic of $C/\mathrm{d.o.f.}=299.69/284$, and no significant residuals remain. 

%%%%%%%%%%%%%%%%%%%%%%%%%%%%%%%%%%%%%%%
\section{Phase-Resolved Variability} %%
%%%%%%%%%%%%%%%%%%%%%%%%%%%%%%%%%%%%%%%
\label{sec:phase-resolved}

\begin{figure}
    \centering
    \includegraphics[width=\columnwidth]{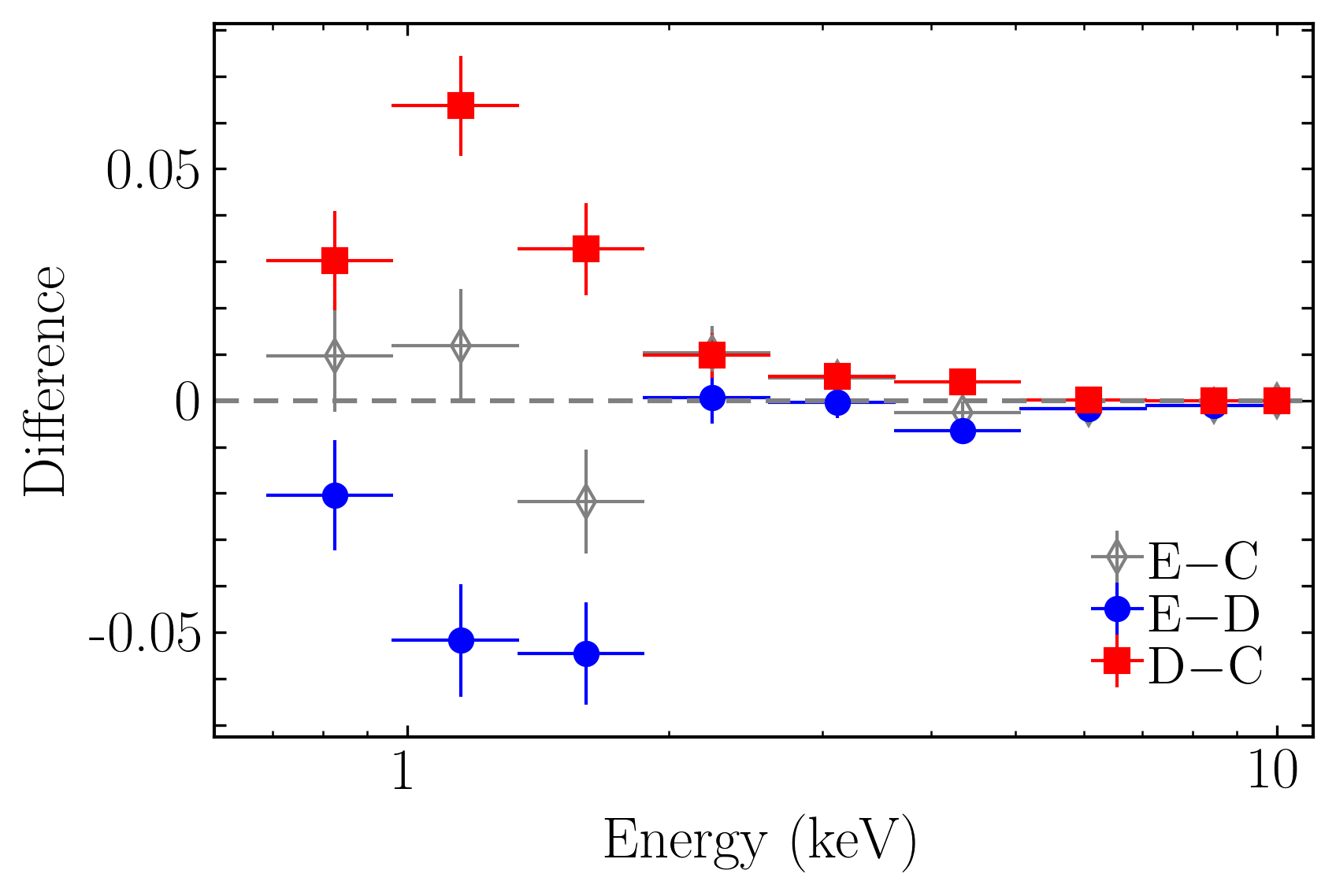}
    \caption{Difference between phase-resolved spectra C, D, and E. Significant differences between segments E and D as well as D and C are present at low energies, while there is little difference between segments E and C.}
    \label{fig:difference}
\end{figure}

In addition to the transition from a soft to hard state evident in the hardness ratio, there is more subtle spectral variability tied to the apparent sinusoidal variations (see Figure \ref{fig:master_lc}). Considering the hard state alone to isolate this mode of variability and to avoid contamination from the soft/hard variability, we considered only spectra C, D, and E for this portion of the analysis. In an attempt to uncover the underlying mechanism driving the oscillatory behaviour, we took the difference of spectra C, D, and E as shown in Figure \ref{fig:difference}. This revealed that these phase-resolved spectra differ only at lower energies, indicating that the oscillatory behaviour in the $1-3$ keV band discovered in Figure \ref{fig:master_lc} is the primary driver of variability during these times. Any changes found between these three spectra is thus associated with the oscillatory behaviour. 

In an attempt to uncover the driver behind this variability, we modelled the C, D, and E phase-resolved spectra, leaving out A and B as they exist in soft and transitional spectral states, respectively. We fit the spectra with the blurred reflection and partial covering models for a variety of parameter combinations. We omitted the soft Comptonisation model from this portion of the analysis as we found it difficult to justify changes in this model over such short time-scales, as variations in the physical parameters of the corona (e.g. accretion rate, electron temperature, optical depth) would be expected to change more slowly than $\sim$40 ks.

Both models could replicate the changes in the spectra, but the difference between measured parameters in spectra C, D, and E are insignificant. No strong statement can be made as to the origin of the variability in the $1-3$ keV band.

%%%%%%%%%%%%%%%%%%%%%%%
\section{Discussion} %%
%%%%%%%%%%%%%%%%%%%%%%%
\label{sec:discussion}

%%%%%%%%%%%%%%%%%%%%%%%%%%%%%%%%%%%%
\subsection{The Fe K$\alpha$ region}

The Fe K$\alpha$ region could be well-fitted with a single narrow Gaussian at $6.4$ keV and a broader Gaussian profile ($\sigma=400\pm200$ eV) at $6.7$ keV.  Broad, ionized lines, presumably originating from the accretion disc have been proposed in some objects (e.g. \citealt{Pounds+2001,Reeves+2001}).  The corresponding full-width at half-maximum of this broad line corresponds to a velocity of approximately $40000$ km s$^{-1}$, which is much greater than the velocities seen in the Balmer lines of Mrk~530 ($\sim 6000$ km s$^{-1}$, \citealt{Kollatschny+2000}). Such a velocity would place the origin of this line between the BH and the broad-line region, likely in the accretion disc.

Alternatively, the Fe K$\alpha$ region can be well-fitted with a complex of narrow features, independent of the continuum model used to fit the broadband spectrum. A $6.4$ keV Fe I K$\alpha$ line and narrow ionised Fe XXV K$\alpha$ and Fe XXVI K$\alpha$ lines at $6.7$ keV and $6.97$ keV, respectively, are able to account for all excess residuals in the Fe K$\alpha$ region with good fit quality (Section \ref{sec:feka}). In this scenario, the origin of these emission lines is beyond the inner accretion disc, as all lines are narrow and have not been relativistically broadened by the effects of the SMBH. These narrow lines thus arise from Compton-thin material (such as the Broad or Narrow Line Region), or Compton-thick material (such as the torus) \citep{Matt+2001}. 

The data are consistent with a Compton shoulder at $6.3$ keV which points to an origin from Compton-thick matter like the torus \citep{Matt+1991,Matt+1996a}. The lines could originate from fluorescence and resonant scattering in a highly ionised outer layer of photoionised matter in the torus \citep{Matt+1996b,BianchiMatt2002,Costantini+2010}. The equivalent width of the lines in Mrk~530 are consistent with originating from photoionised material \citep{BianchiMatt2002}. Though the data do not require inclusion of the Compton shoulder, the torus as the origin of the narrow lines is physically realistic as Mrk $530$ is classified as a Seyfert $1.5$, so we are observing the system at high inclination and seeing much of the torus. The scenario is consistent with observations of other sources like NGC 5506, an intermediate Seyfert with a complex Fe region \citep{BianchiMatt2002}, and Mrk 279, a Seyfert 1 with ionised lines likely originating from the tours \citep{Costantini+2010}. A future mission like XARM, with Hitomi-like \citep{Takahashi+2016} spectral resolution will advance this work.

%%%%%%%%%%%%%%%%%%%%%%%%%%%%%%%%%%%%%%%%%%%%%%%%%%%%%%%%
\subsection{The broadband X-ray spectrum and variability}

The X-ray spectrum of Mrk~530 transitions smoothly from a power law to a soft excess at about $E<2$~keV.  The iron line complex exhibits no strong evidence of a relativistically broadened Fe~K$\alpha$ emission line and any hard excess emission above $10$ keV that could be associated with a Compton hump is also weak. As such, the source spectrum is reminiscent of complicated sources like Akn 120 (e.g. \citealt{Matt+2014}).

The X-ray continuum of Mrk~530 could be well-fitted with a number of physical models like blurred reflection, partial covering, and soft Comptonisation.  None of the models could be distinguished by quality of fit.

In the partial covering and Comptonisation scenarios, the bulk of the emission above 10 keV is fitted with the {\sc pexmon} component that is used to mimic distant reflection from the cold torus.  In the blurred reflection scenario, the {\sc pexmon} component is weaker as the blurred reflector contributes to some of the high-energy emission. 

However, the reflection fraction in the blurred reflector model is below unity ($R\approx0.22$), which is consistent with anisotropic emission away from the accretion disc (e.g. \citealt{Gonzalez+2018}). This could be beamed emission that arises in the base of a jet (e.g. \citealt{Gallo+2015,WilkinsGallo2015b}), but radio studies of Mrk 530 (e.g. \citealt{Lal+2011}) have not revealed evidence of a jet component.  The need for a shallow emissivity profile in the X-ray model also indicates that any point-like corona would be located at relatively large distances from the black hole and not strongly influenced by light bending effects.

During the course of this \textit{Suzaku} observation, Mrk 530 exhibited two distinct spectral states: a soft state which smoothly transitioned to a hard state. The variability between these two states can be well-described by a blurred reflection model with changes in photon index and reflection fraction between the two states (Section \ref{sec:reflection_hardsoft}). The photon index is greater (steeper) for the soft state, and lower (flatter) for the hard state. The change in photon index indicates changes in the corona (e.g. temperature, density), or could be attributed to changes in velocity if the source is indeed moving away from the disc (e.g. \citealt{Gonzalez+2018}), which would also result in changes in the reflection fraction. 

The spectral state change can also be explained through a partial covering scenario with changes in covering fraction of an ionised absorber (Section \ref{sec:pc_hardsoft}). Here the hard state sees an increase in the covering fraction which attenuates low-energy emission, effectively hardening the spectrum. The covering fraction itself can represent a single patchy cloud or numerous individual clouds. Therefore, a change in covering fraction could also indicate a change in the number of clouds along the line of sight.

Alternatively, the spectral state change could be explained with a soft Comptonisation model that changes in photon index and Eddington ratio (Section \ref{sec:compt_hardsoft}). As was the case for the blurred reflection model, the photon index was found to be greater for the soft state and lower for the hard state, indicating changes in the primary emitter between the states. 

This model produced a lower Eddington ratio along with a steeper power-law during the soft state, exhibiting the softer-when-brighter behaviour frequently observed in Seyfert galaxies (e.g. \citealt{Shemmer+2006,SobolewskaPapadakis2009}).   However, understanding the rapid changes in the Eddington ratio is difficult.

It is plausible that the spectral variability in Mrk 530 is not due to simply one of these examined models, but rather some combination of them (e.g. \citealt{Parker+2018}).

%%%%%%%%%%%%%%%%%%%%%%%%%%%%%%%%%%%%%%%%%
\subsection{Light curve variability}
The light curve displays variations that resemble two cycles of a sinusoidal function. Though a sinusoid fit to the data is not statistically significant ($\chi^2_\nu=2.57$), the oscillations have a period of $T \approx 86$ ks, which for a BH mass of $1.15\times10^8$ M$_{\odot}$ \citep{Woo+2012}, is consistent with the dynamical timescale at $\sim10$ $r_g$ in a standard accretion disc. The radius is completely consistent with where we expect X-ray emission to be coming from. However, it has been shown that ordinary red noise can commonly mimic a few cycles of sinusoidal behaviour \citep{Vaughan+2016}. In cases of robust detection of periodicity, the QPO behaviour has been detected in a narrow band in addition to broadband aperiodic variability. For example, the QPO behaviour in RE J1034+396 does not dominate the red noise variability, but appears in addition to it \citep{Gierlinski+2008}.

\cite{HalpernMarshall1996} observed the Seyfert 1 galaxy RX J$0437.4-4711$ with the \textit{EUVE} satellite and found the object to exhibit periodic behaviour in the soft X-ray regime ($0.12-0.18$ keV) with a period of $\sim0.9$ days, which they note could correspond to an orbital timescale around such a SMBH. \cite{Halpern+2003} found this detection to be significant on the $>95$ per cent level.

To ensure that the field around Mrk 530 was free of contamination from variable sources, the surrounding field of view was checked for other objects that could be contributing to this periodic detection. Two X-ray sources that lay within the source extraction radius were found, 2XMM J$231856.1+001403$ and 2XMM J$231857.3+001341$. However, the flux of these sources was found to be negligible ( $5.5\times10^{-13}$ erg cm$^{-2}$ s$^{-1}$ and $1.1\times10^{-13}$ erg cm$^{-2}$ s$^{-1}$ in the $0.2-12$ keV band) compared to the flux of Mrk 530 ($4.3\times10^{-11}$ erg cm$^{-2}$ s$^{-1}$ in the same band), and so these sources can be confidently discounted as the source of variability in this observation.

%%%%%%%%%%%%%%%%%%%%%%%%
\section{Conclusions} %%
%%%%%%%%%%%%%%%%%%%%%%%%
\label{sec:conclusions}

Mrk 530 was observed by \textit{Suzaku} and found to exhibit two distinct types of variability during the course of the observation: a spectral state change which transitioned smoothly from a soft state to a hard state, and apparent sinusoidal variability in a narrow energy band.

To characterise the average spectrum of the observation as well as the spectral changes between the soft and hard states, three different physical models were fit to the data: a blurred reflection model, a partial covering model, and a soft Comptonisation model. All three models were able to describe the average spectrum and changes between the two spectral states reasonably well. Fits to the Fe~K$\alpha$ band favour multiple narrow emission lines likely originating from neutral and ionised material in the distant torus.

The apparent periodicity in the light curve is statistically insignificant. We were interested in the driver behind this variability given that the period corresponds to the orbital timescale of the BH at a radius of $\sim10$ $r_g$. However, we were unable to constrain the driver of this variability with current data quality.

Deeper observations over a broader band will provide the opportunity to distinguish various continuum models and further investigate continuum flux variability.

%%%%%%%%%%%%%%%%%%%%%%%%%%%%%%%%%%%%%%%%%%%%%%%%%%
\section*{Acknowledgements}
This work made use of data from the \textit{Suzaku} satellite, a collaborative mission between the space agencies of Japan (JAXA) and the USA (NASA). We thank referee for their careful reading and helpful comments on the original manuscript.

%%%%%%%%%%%%%%%%%%%%%%%%%%%%%%%%%%%%%%%%%%%%%%%%%%

\bibliographystyle{mnras}
\bibliography{m530_ref.bib}
%%%%%%%%%%%%%%%%%%%%%%%%%%%%%%%%%%%%%%%%%%%%%%%%%%

% Don't change these lines
\bsp	% typesetting comment
\label{lastpage}
\end{document}